\documentclass{PoS}

\title{Semi-annihilation of Dark Matter }

\ShortTitle{Semi-annihilation of Dark Matter }

\author{\speaker{Francesco D'ERAMO}%
        \\
       Massachusetts Institute of Technology\\
       E-mail: \email{fderamo@mit.edu}}

\abstract{The semi-annihilation reaction takes the schematic form $\psi_i \psi_j \rightarrow \psi_k \phi$, where $\psi_i$ are stable dark matter particles and $\phi$ is an unstable state. Such reactions are allowed when dark matter is stabilized by a larger symmetry than just $Z_2$. They lead to non-trivial dark matter dynamics in the early universe,  and the thermal production of the relic particles can be completely controlled by semi-annihilations. This process might also take place today in the Milky Way, enriching the (semi-)annihilation final state spectrum observed in indirect detection experiments.}

\FullConference{Identification of Dark Matter 2010\\
		 July 26 - 30 2010\\
		 University of Montpellier 2, Montpellier, France}

\begin{document}

\section{Introducing semi-annihilation}
The evidence for dark matter is one of the best motivations for physics beyond the Standard Model (SM) \cite{Bertone:2004pz}. A well-motivated dark matter candidate are so-called Weakly Interacting Massive Particles (WIMPs), whose production in the early universe is thermal \cite{Lee:1977ua, Kolb:1990vq}. The thermal freeze-out is very predictive, thus it is important to know how to correctly compute the relic density, especially when the relic computation cannot be reduced to the standard case (see e.g. \cite{Griest:1990kh}).  

\begin{figure}
$\begin{array}{c@{\hspace{1in}}c}
\includegraphics[scale=0.55]{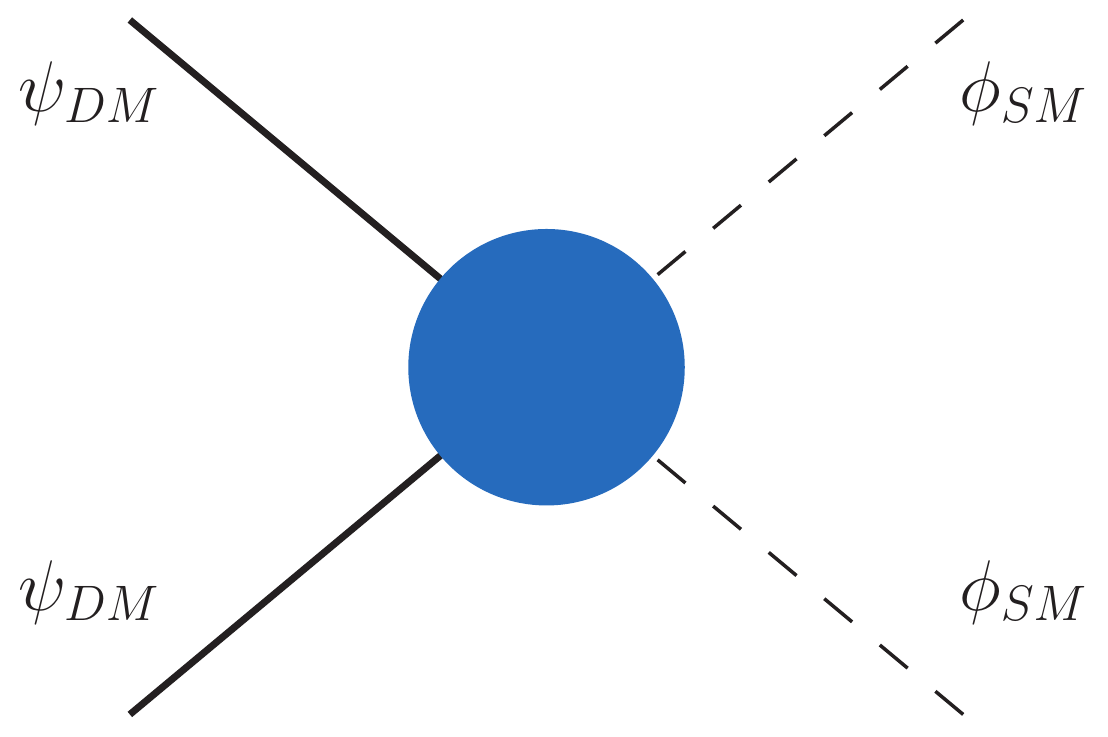} &
\includegraphics[scale=0.55]{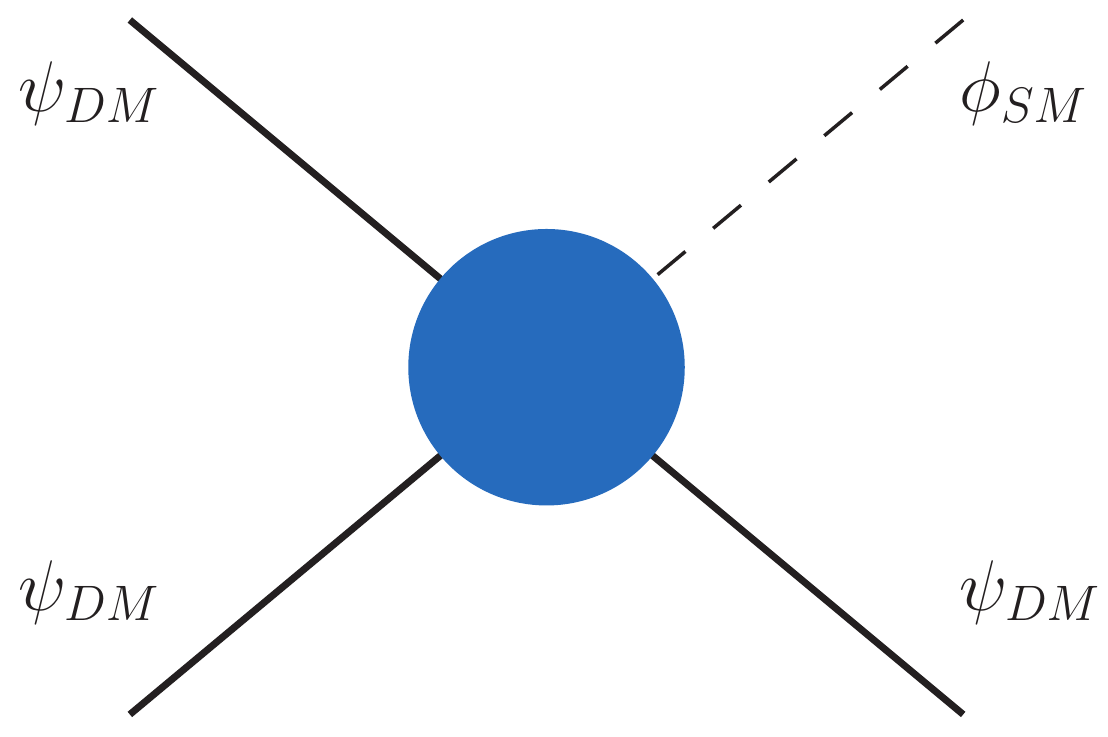} \\ [0.1cm]
\mbox{\bf (a)} & \mbox{\bf (b)}
\end{array}$
\caption{Reactions that keep the dark matter particles in thermal equilibrium in the early universe: (a) annihilation; (b) semi-annihilation. The field $\phi_{SM}$ decays to SM states.}
\label{fig:AnnVsSemi}
\end{figure}

The thermal abundance can be dramatically affected by the presence of the ``semi-annihilation'' reaction. If the dark matter is composed of more than one stable component $\psi_i$, it takes the form
\begin{equation}
\psi_i \psi_j \rightarrow \psi_k \phi ,
\end{equation}
where $\phi$ is a SM state or a new particle which decays to the SM. We see that unlike ordinary annihilation where the total dark matter number changes by two units, in semi-annihilation the total dark matter number changes by only one unit, as shown in figure \ref{fig:AnnVsSemi}.  As long as the triangle inequality $m_k < m_i+m_j$ is satisfied (as well as its crossed versions), the semi-annihilation reaction $\psi_i \psi_j \rightarrow \psi_k \phi$ can take place without making any relic particle unstable. Such reactions take place once we relax some assumptions about the symmetry structure of WIMP interactions. For example the dark matter can be stabilized by a $Z_3$ symmetry \cite{Agashe:2004ci},  composed of non-Abelian gauge bosons \cite{Hambye:2008bq}, or more generally stabilized by  ``baryon'' and/or ``flavor'' symmetries, as in QCD-like theories. We study two explicit toy examples of such models \cite{D'Eramo:2010ep}. We find that to correctly compute the relic abundance, semi-annihilation must be included, and it can dominate over standard annihilation. We also explore the effects of semi-annihilations on indirect detection experiments.

We introduce semi-annihilation with a simple case.  We assume dark matter to be composed of a complex scalar $\chi$, which is stabilized by a $Z_3$ symmetry. The $\chi$ particle interacts with a real scalar $\phi$, which eventually decays to SM states. The $\chi$ particles are kept in thermal equilibrium in the early universe by the reactions $\chi \bar{\chi} \rightarrow \phi \phi$ and $\chi \chi \rightarrow \bar{\chi} \phi$. While we focus on the case of a field $\phi$ ``portal'' to the SM  \cite{Pospelov:2007mp}, in more general dark matter scenarios $\phi$ could be a SM field itself \cite{Burgess:2000yq}.

Single-component dark matter with a $Z_3$ symmetry is not representative of models in which semi-annihilation is relevant. The second model we consider is a minimal multi-component dark matter model having semi-annihilations among its allowed reactions.\footnote{A more realistic construction, where a supersymmetric gauge theory with $N_f = N_c + 1$ is considered, can be found in \cite{D'Eramo:2010ep}.  Here, we prefer to focus on a simpler model avoiding unnecessary complications caused by having superpartners.} In this toy system, dark matter is composed of two stable components, a complex scalar field $\chi$ and a vector-like fermion $b$ and $b^c$. We impose a $U(1)$ global symmetry under which $b$, $b^c$, and $\chi$ have charges $+1$, $-1$, and $-2$, respectively. We also introduce a real scalar portal $\phi$ field, which we assume to be in thermal equilibrium in our calculation and decays to SM states. Fermion number conservation always guarantees that $b$ is stable, whereas stability for $\chi$ requires the triangle inequality rule $m_\chi \leq 2 m_b$ (isosceles triangle). The model and its symmetries are summarized in table \ref{tab:bbchisym}.

\begin{table}
\begin{center}
\begin{tabular}{|c|c|c|}
\hline
Fields & Spin & $U(1)$ charge  \\ 
\hline\hline
$b$ & Weyl left & $+1$ \\
$b^c$ & Weyl left & $-1$  \\
$\chi$ & complex scalar & $-2$  \\
$\phi$ & real scalar & $0$  \\ 
\hline
\end{tabular}
\end{center}
\caption{Field content and symmetries of a minimal multi-component model with semi-annihilation.}
\label{tab:bbchisym}
\end{table}

The remainder is structured as follows.  In Section \ref{sec:relic} we present a study of the relic density in the two toy models we consider. In Section \ref{sec:indirect} we explore the effects of semi-annihilations on indirect detection experiments, and conclude in Section \ref{sec:con}. 

\section{Implications for the relic density}
\label{sec:relic}
The number density evolution of $\chi$ in the model with $Z_3$ symmetry is described by \footnote{We correct the Boltzmann equation in \cite{D'Eramo:2010ep} with the $1/2$ factor in the semi-annihilation term. We thank Brian Batell.}
\begin{equation}
\frac{d n_{\chi}}{d t} + 3 H n_{\chi} = - \langle \sigma v \rangle_{\chi\bar{\chi}\rightarrow\phi\phi} \left[n_{\chi}^2 - n_{\chi}^{{\tt eq}\, 2}\right] 
- \frac{1}{2} \langle \sigma v \rangle_{\chi\chi\rightarrow\bar{\chi}\phi} \left[n_{\chi}^2 - n_{\chi} n_{\chi}^{{\tt eq}}\right].
\label{eq:z3boltzmann}
\end{equation}
In the $\langle \sigma v \rangle_{\chi\chi\rightarrow\bar{\chi}\phi}=0$ limit we recover the standard case.  The above Boltzmann equation can be solved semi-analytically in analogy with the method in \cite{Kolb:1990vq}, by solving the equation in two regimes, early and late times, and then matching the two solutions at the freeze-out point.  The freeze-out temperature $x_f$, where we define $x = m_{\chi}/T$, is found by solving the equation
\begin{equation}
x_f = \log\left[0.038\,c(c+2)\langle \sigma v \rangle_{\chi\bar{\chi}\rightarrow\phi\phi} \frac{g_{\chi}\,m_{\chi}\,M_{\tt Pl}}{\sqrt{g_{*}\,x_f}}\right] + 
\log\left[1 + \frac{1}{2}\,\frac{c+1}{c+2}\,\frac{\langle \sigma v \rangle_{\chi\chi\rightarrow\bar{\chi}\phi}}{\langle \sigma v \rangle_{\chi\bar{\chi}\rightarrow\phi\phi}}\right]
\label{eq:xfZ3},
\end{equation}
where $c = \sqrt{2}-1$ for s-wave amplitudes \cite{Kolb:1990vq}. The first term on the right-hand side of (\ref{eq:xfZ3}) is the solution we usually have in the Lee-Weinberg scenario.  Thus we see that the effect of semi-annihilation is to shift the freeze-out temperature by only a small logarithmic amount. The mass density of the $\chi$ particle today is given by \begin{equation}
\Omega_{\chi} h^2 = 2 \times \frac{1.07 \times 10^9 \, {\rm GeV^{-1}}}{\sqrt{g_*} M_{{\tt Pl}} J(x_f)},\qquad\qquad J(x_f) = \int_{x_f}^{\infty} \frac{dx}{x^2}
\left[\langle \sigma v \rangle_{\chi\bar{\chi}\rightarrow\phi\phi} + \frac{1}{2} \langle \sigma v \rangle_{\chi\chi\rightarrow\bar{\chi}\phi}\right] .
\label{eq:Omegah2Z3}
\end{equation}
In single component dark matter models, the conventional assumption is that we have only annihilation (namely $\langle \sigma v \rangle_{\chi\chi\rightarrow\bar{\chi}\phi} = 0$), thus once we fix the dark matter mass $m_{\chi}$ the cross section $\langle \sigma v \rangle_{\chi\bar{\chi}\rightarrow\phi\phi}$ is uniquely determined by the relic density measurement. However, if we allow also semi-annihilations, a wider region of parameter space gives the correct relic density, thus the thermal production of the relic particles can also be completely controlled by semi-annihilations.

In models which involve more than one stable dark matter component a semi-analytical solution for the relic density is simply not possible, and one must resort to numerically solving the complete set of coupled Boltzmann equations.  In particular, the simplifying assumptions made when analyzing co-annihilating models \cite{Griest:1990kh} are not applicable for semi-annihilation \cite{D'Eramo:2010ep}.  As a consequence, the dark matter dynamics in the presence of semi-annihilation are far more varied than in decoupled multi-component scenarios. 

In our second example model, the relic abundance of $b$ and $\chi$ is described by a system of two coupled Boltzmann equations.  We can now write down and solve numerically the Boltzmann equations for the relic abundance of $b$ and $\chi$ and see how semi-annihilations affect the dark matter relic density. Once we fix the dark matter masses and limit our consideration to $s$-wave annihilation, there are four free parameters, which we choose to be the following $s$-wave matrix elements
\begin{equation}
\mathcal{M}_{b \bar{b} \rightarrow \phi \phi} = \alpha, \qquad \mathcal{M}_{\chi \bar{\chi} \rightarrow \phi \phi} = \beta, 
\qquad \mathcal{M}_{b \bar{b} \rightarrow \chi \bar{\chi}} = \kappa,  \qquad \mathcal{M}_{\chi b \rightarrow \phi \bar{b}} = \epsilon.
\label{eq:bbchireacs}
\end{equation}
The first two amplitudes come from the standard annihilation to light particles in the final state, the third from the process where two dark matter particles of one species are entirely converted to two dark matter particles of the other species. This process is phase space suppressed for typical kinetic energies at freeze-out. The last corresponds to the semi-annihilation process, and we are interested to see how the relic abundance changes with the parameter $\epsilon$.  

\begin{figure}
$\begin{array}{ccc}
\includegraphics[scale=0.45]{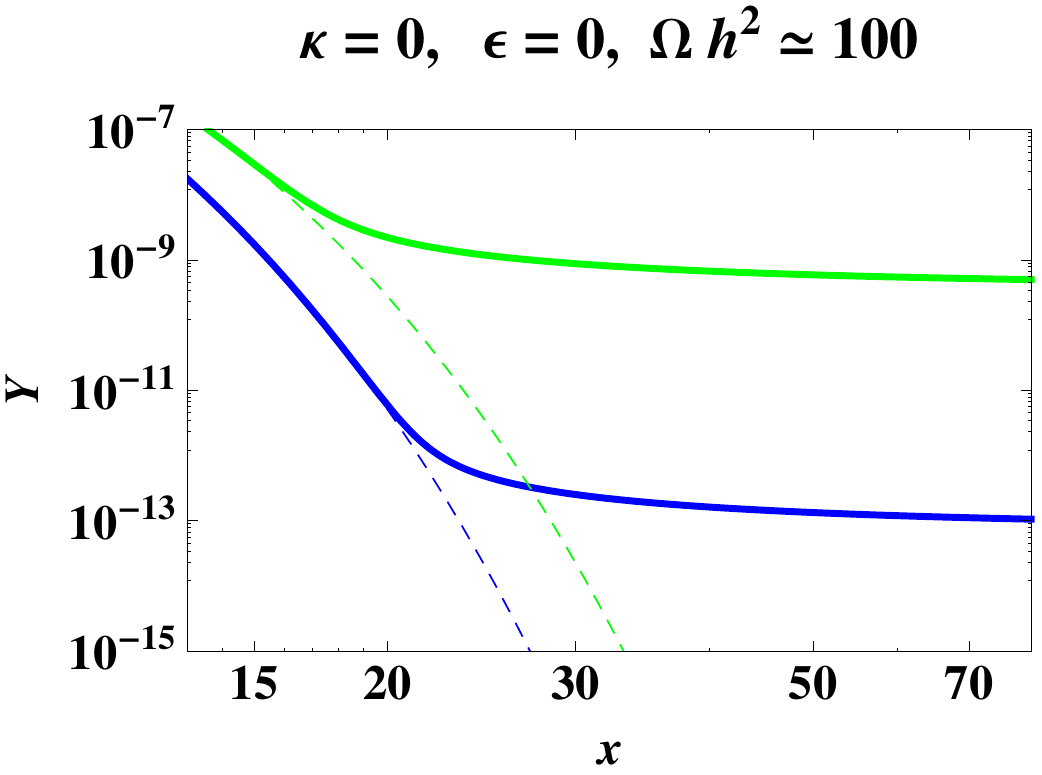} & 
\includegraphics[scale=0.45]{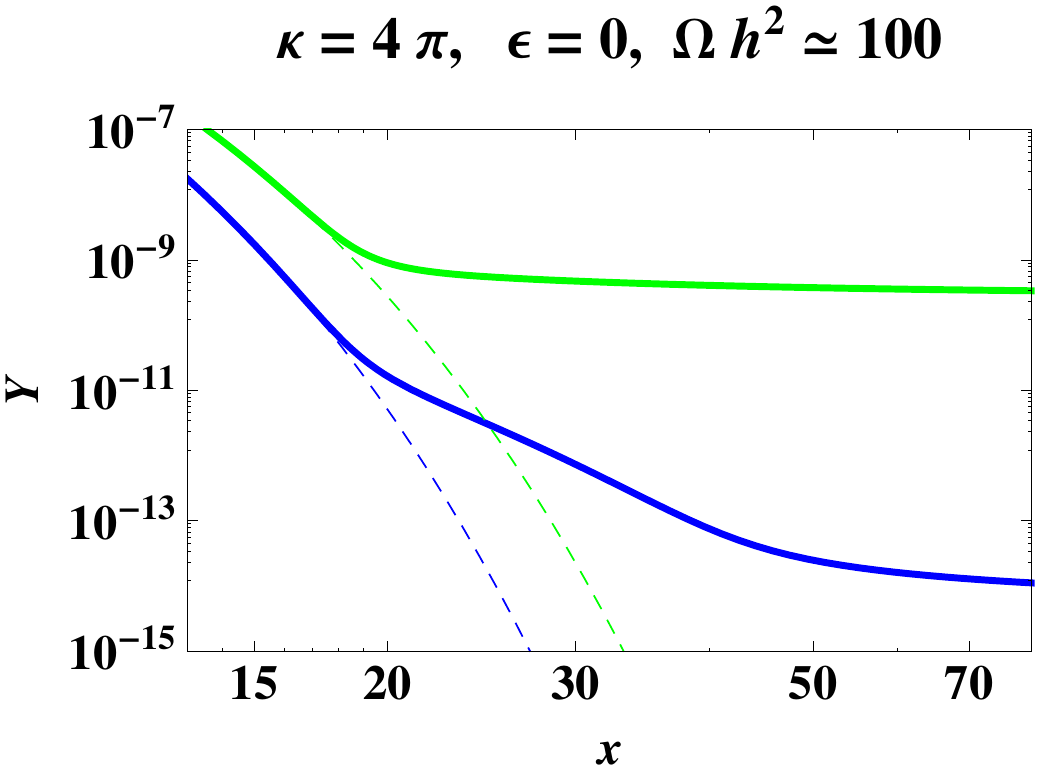} &
\includegraphics[scale=0.45]{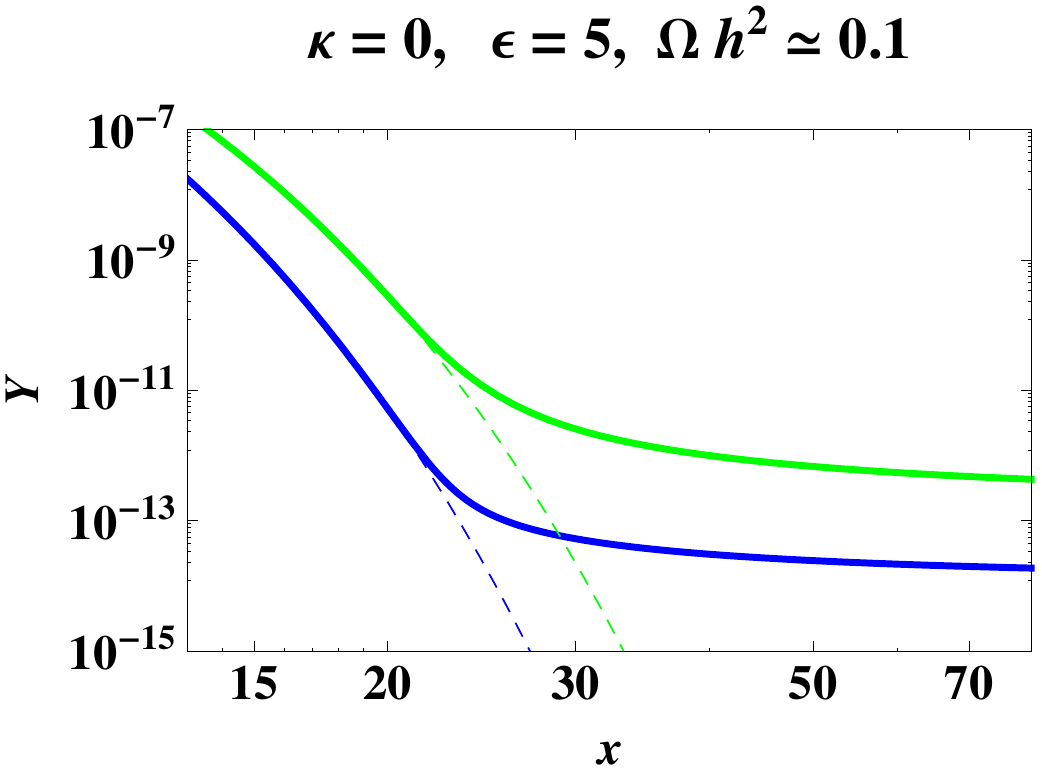}  \\
\mbox{\bf (a)} & \mbox{\bf (b)} & \mbox{\bf (c)}  \\ 
\end{array}$
\caption{$Y$ vs. $x$ evolution in the $bb\chi$ model for the $b$ (blue lines) and the $\chi$ (green lines). We define $Y_i = n_i/s$, where $i=b, \chi$ and $s$ is the entropy density. We plot the equilibrium values (dashed lines) and the numerical solutions (solid lines). We fix $m_\chi = 0.8 \,{\rm TeV}$, $m_b = \, 1 {\rm TeV}$, $\alpha = 2$ and $\beta = 0.01$. The values of $\kappa$, $\epsilon$ and $\Omega_{{\rm DM}} h^2$ are shown in each plot. Here, $\chi$ particles would be over produced (a), but because of phase space suppression $\kappa$ ($b \leftrightarrow \chi$ conversion) is ineffective (b). However, $\epsilon$ (semi-annihilation) is never phase space suppressed and can restore the desired relic density (c).}
\label{fig:bbchi3}
\end{figure}

In our numerical study we fix the dark matter masses to be $m_\chi = 0.8 \, {\rm TeV}$ and $m_b = 1 \, {\rm TeV}$ for concreteness, and we vary the matrix elements amplitudes. We fix the values of $\alpha$ and $\beta$ (annihilation amplitudes) and study how the relic density is affected once we turn on the parameters $\kappa$ ($b \leftrightarrow \chi$ conversion) or $\epsilon$ (semi-annihilation). The main difference between semi-annihilation and species conversion is that the former is never phase space suppressed. This is beautifully illustrated in the case when pure annihilations would give a lot of $\chi$ and very few $b$, as shown in figure \ref{fig:bbchi3}({\rm a}) for $\alpha=2$ and $\beta=0.01$. The semi-annihilation process $\chi b \rightarrow \bar{b} \phi$ is still able to destroy $\chi$ particles and bring the relic density back to the observed value, figure \ref{fig:bbchi3}({\rm c}) for $\epsilon = 5$. However, the reaction $\chi\bar{\chi} \rightarrow b \bar{b}$ is powerless, even if we badly break perturbation theory for $\kappa = 4 \pi$ as in figure \ref{fig:bbchi3}({\rm b}). This reaction is forbidden at zero kinetic energy, and at freeze-out the thermal energy is very small compared with the mass splitting between the two particles. Thus, semi-annihilation is a truly unique species changing interaction that affects early universe cosmology.

\section{Implications for indirect detection}
\label{sec:indirect}
We now explore the implications of semi-annihilations for dark matter detection experiments. Since semi-annihilation does not give any new contributions to direct detection rates, we focus on indirect detection only. There has been much recent interest on indirect detection of dark matter, motivated by observations suggesting the presence of a new primary source of galactic electrons and positrons \cite{Adriani:2008zr}. These anomalies might be evidence for dark matter annihilation or decay in our galaxy.

\begin{figure}
$\begin{array}{c@{\hspace{1in}}c}
\includegraphics[scale=0.46]{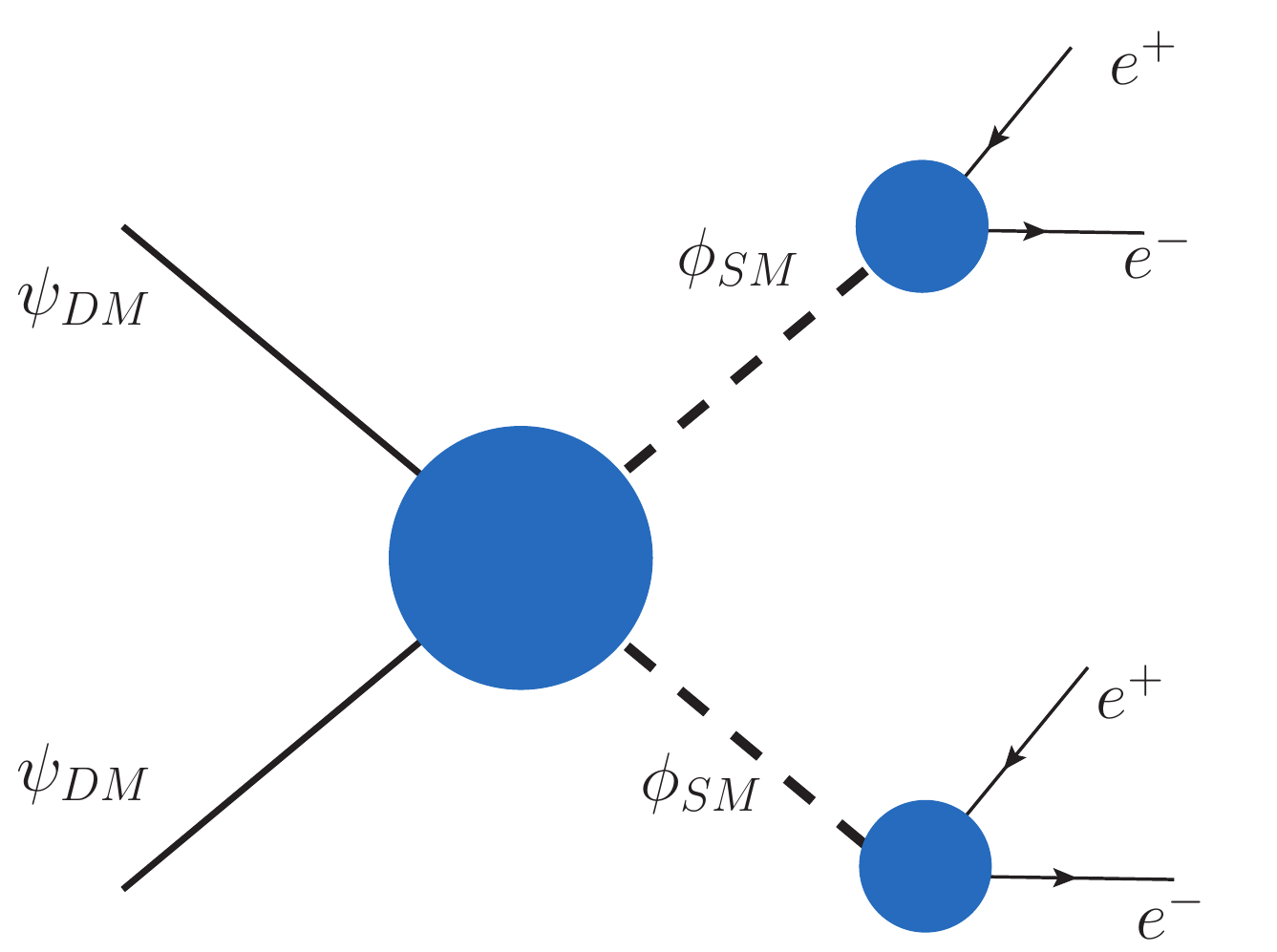} &
\includegraphics[scale=0.46]{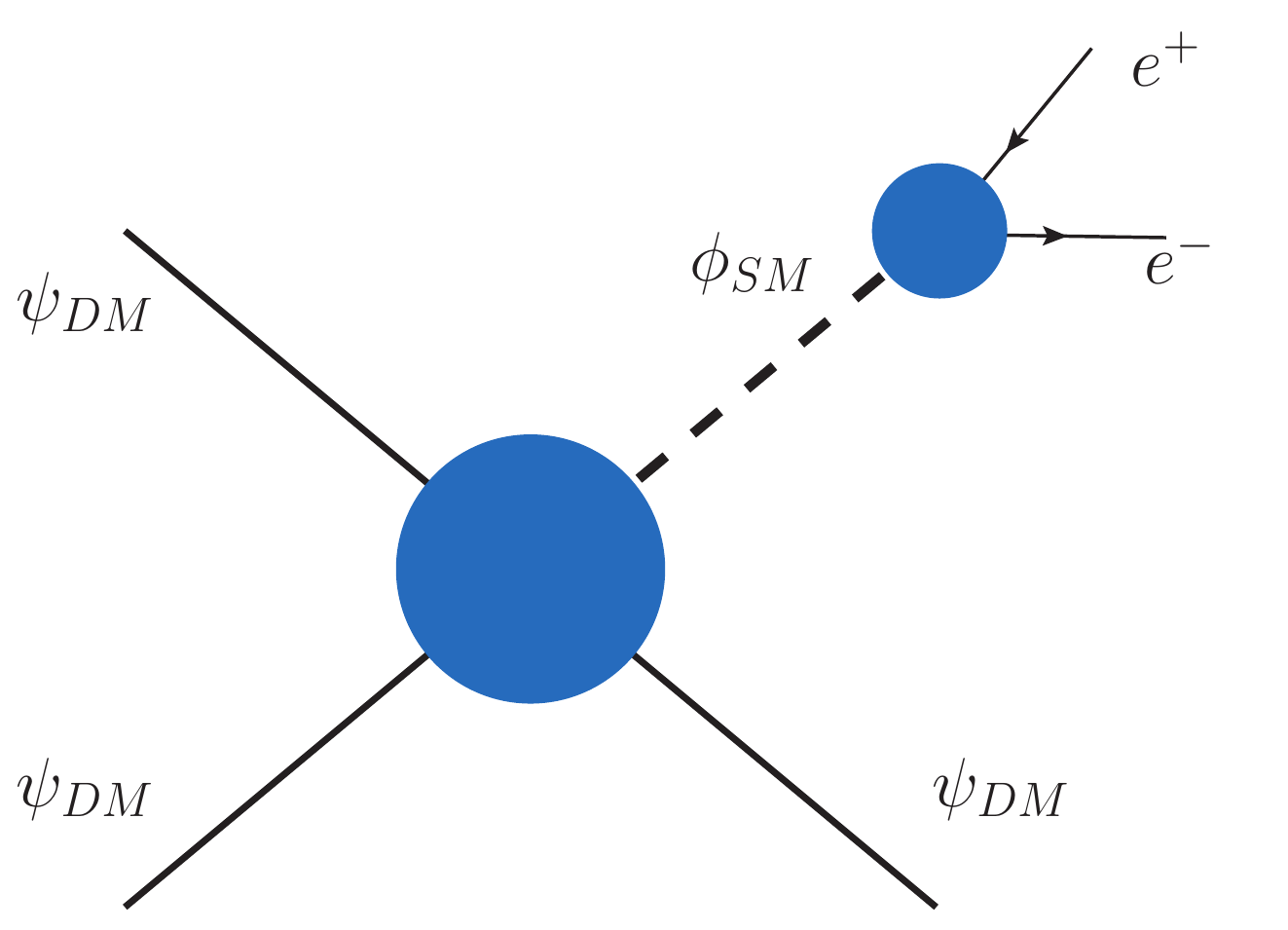} \\ [0.1cm]
\mbox{\bf (a)} & \mbox{\bf (b)}
\end{array}$
\caption{Two different contributions to indirect detection spectra: (a) annihilation; (b) semi-annihilation. The field $\phi_{SM}$ is a portal field, but in more general dark matter scenarios $\phi$ could be a SM field itself.}
\label{fig:IndirectAnnVsSemi}
\end{figure}

In general, semi-annihilation does not lead to the ``boost factor'' necessary to explain the magnitude of these candidate dark matter indirect signals. However the presence of such a reaction can still have considerable implications for indirect searches. Semi-annihilations are additional channels to produce light particles in dark matter interactions in our galaxy, as shown in figure \ref{fig:IndirectAnnVsSemi}, which have no analog in standard multi-component scenarios. Thus the predicted spectrum is enriched with respect to the standard scenarios. In addition, our examples models can account for the fact that dark matter (semi-)annihilates preferably into leptons; if we assume a portal $\phi$ field mass $m_{\phi} \leq 2 \, {\rm GeV}$ the decay of $\phi$ to antiprotons is kinematically forbidden \cite{Pospelov:2007mp}.

\begin{figure}
$\begin{array}{cc}
\includegraphics[scale=0.65]{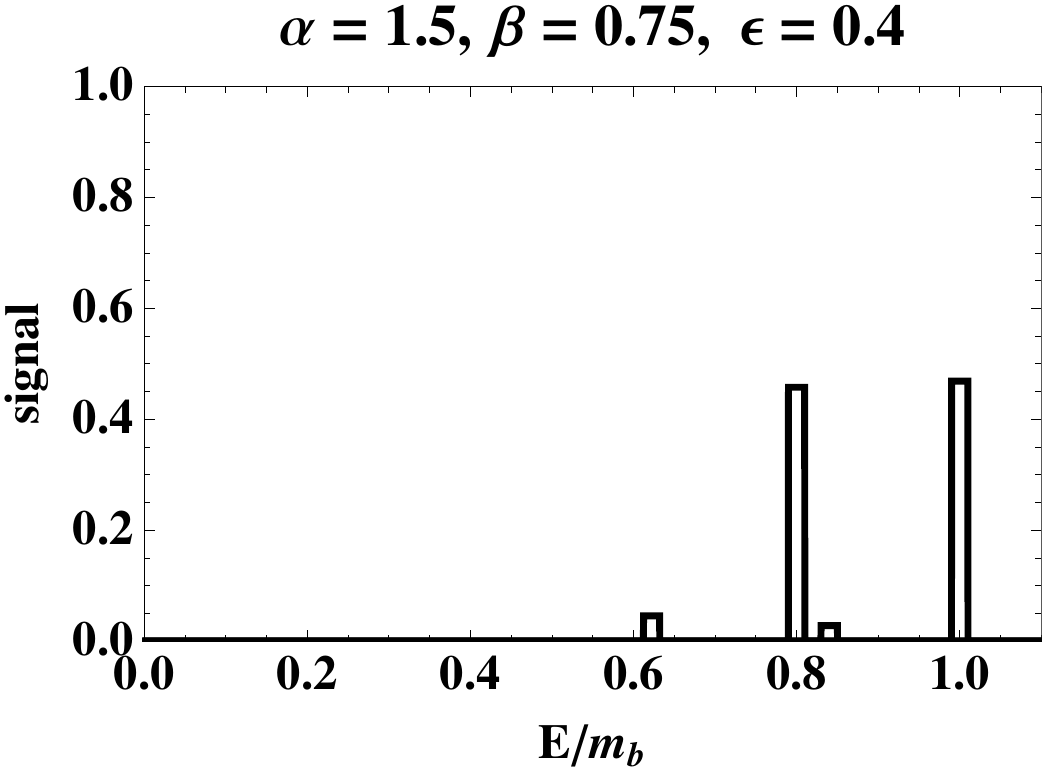} & \hspace{0.5cm}
\includegraphics[scale=0.65]{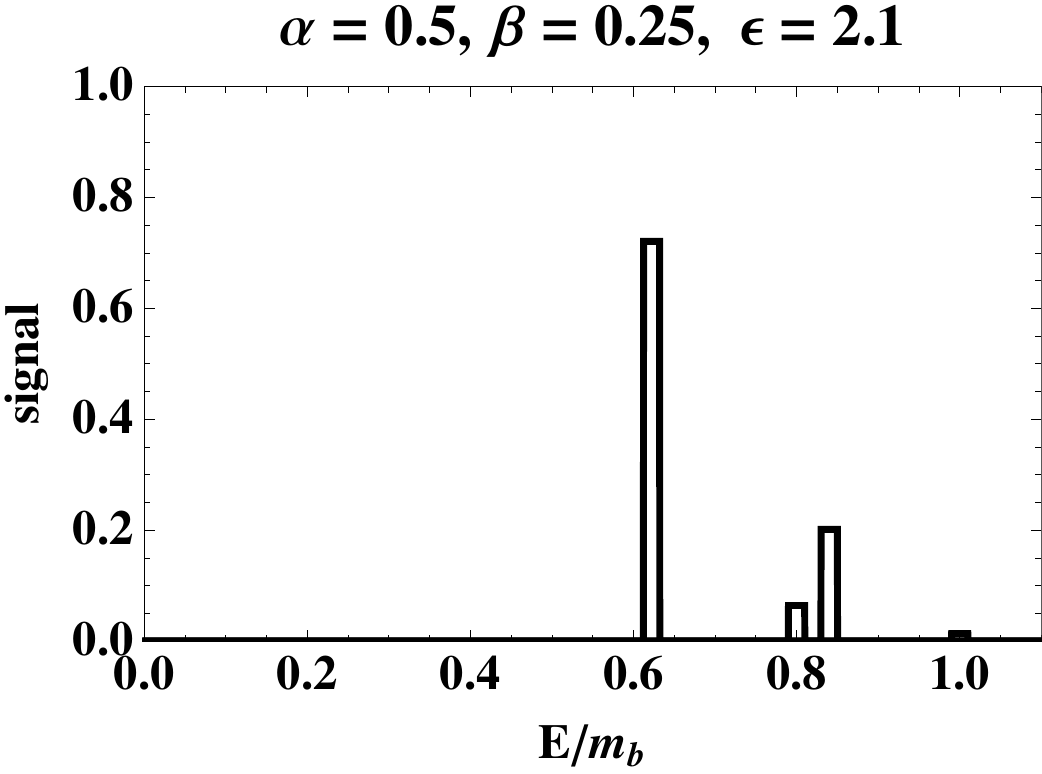} \\
\mbox{\bf (a)} & \mbox{\bf (b)} 
\end{array}$
\caption{Spectra of $\phi$ particles in the $bb\chi$ model for thermal production mostly controlled by: ({\rm a}) annihilations, ({\rm b}) semi-annihilations. The values of the matrix element amplitudes are shown in each plot.}
\label{fig:bbchispec}
\end{figure}

We present here the results for the $b b \chi$ model.\footnote{For the spectra in the $Z_3$ model see reference \cite{D'Eramo:2010ep}.} In this case, there are four reactions to produce $\phi$ particles in the final state, the ordinary annihilations of the two components and two semi-annihilations. Thus the spectrum is this multi-component case is quite rich, since the four reactions yeld four monochromatic $\phi$ lines. We present numerical results for the $\phi$ spectrum for the same values of the dark matter masses chosen in Section \ref{sec:relic}, namely $m_\chi = 0.8 {\rm TeV}$ and $m_b = 1 {\rm TeV}$. To simplify the following discussion we perform our analysis for $\kappa=0$. Two example spectra are shown in figure \ref{fig:bbchispec}, where two opposite cases are considered. In the first case, figure \ref{fig:bbchispec}({\rm a}), the semi-annihilation contribution is taken to be very small, thus the pure annihilation lines dominate. When the thermal production is mostly controlled by semi-annihilations, as in figure \ref{fig:bbchispec}({\rm b}), the semi-annihilation lines overwhelm the standard contributions.

One expects the indirect detection spectra in models with more than two stable components to have an even richer structure of $\phi$ lines. Of course, once convolved with the $\phi$ decays, the spectrum is less distinct \cite{D'Eramo:2010ep}. More detailed study is necessary to know whether such a spectrum can be measured in $\gamma$ rays telescopes.

\section{Conclusions}
\label{sec:con}

We relaxed some assumptions about the symmetry structure of WIMP interactions by allowing relic particles to ``semi-annihilate''.  We studied two explicit examples where semi-annihilation is present:  a single species dark matter model with a $Z_3$ symmetry and a multiple species dark matter model with ``baryon'' and ``flavor'' symmetries. We saw that the semi-annihilation process can have a considerable effect on the dark matter relic abundance, and the dark matter dynamics in the presence of semi-annihilation are far more varied than in decoupled multi-component scenarios.

The inclusion of semi-annihilation has interesting implications for indirect detection experiments.  The overall integrated flux is not very affected by semi-annihilation.  However, the final state spectrum in semi-annihilating models is far richer than in standards scenarios because of the differing kinematics between semi-annihilation and ordinary annihilation.

In the language of \cite{Griest:1990kh}, semi-annihilation is in some ways the ``fourth exception'' in the calculation of dark matter thermal relic abundances.  We find it intriguing that unlike the traditional three exceptions (co-annihilation, annihilation below a mass threshold, and annihilation near a pole), this fourth exception not only affects dark matter interactions in the early universe, but also leaves an imprint today via the indirect detection spectrum.  We expect that there are a wide variety of multi-component dark matter models with species changing interactions, motivating further studies of the semi-annihilation process.

\section*{Acknowledgments}
We thank Jesse Thaler for collaboration. This work is supported by U.S. Department of Energy (D.O.E.) under cooperative research agreement DE-FG0205ER41360.

\end{document}